\documentclass[letterpaper, 10 pt, conference]{ieeeconf}  

\IEEEoverridecommandlockouts                              

\usepackage[usenames, dvipsnames]{xcolor}
\usepackage[caption=false]{subfig}
\usepackage[]{graphicx} 
\usepackage{booktabs}
\usepackage{tikz}
\usepackage{comment}
\usetikzlibrary{shapes,positioning,arrows}

\definecolor{yes}{RGB}{92,184,92}
\definecolor{no}{RGB}{56,148,240}
\definecolor{error}{RGB}{217,83,79}

\colorlet{mylinkcolor}{BrickRed}
\colorlet{mycitecolor}{Green}
\colorlet{myurlcolor}{NavyBlue}

\usepackage{svg}
\usepackage{subfig}
\usepackage{graphicx}

\makeatletter
\let\NAT@parse\undefined
\makeatother
\usepackage{hyperref}
\hypersetup{
  linkcolor  = mylinkcolor,
  citecolor  = mycitecolor,
  urlcolor   = myurlcolor,
  colorlinks = true,
}
\usepackage[capitalize]{cleveref}

\title{\LARGE \bf
Eyes on the Game: Deciphering Implicit Human Signals\\ to Infer Human Proficiency, Trust, and Intent}

\author{%
Nikhil Hulle\textsuperscript{1,*}, 
St\'ephane Aroca-Ouellette\textsuperscript{1,*}, 
Anthony J. Ries\textsuperscript{2}, 
Jake Brawer\textsuperscript{1},\\
Katharina von der Wense\textsuperscript{1}, 
Alessandro Roncone\textsuperscript{1}%
\thanks{\textsuperscript{*}Equal Contribution.}%
\thanks{\textsuperscript{1}Department of Computer Science, University of Colorado Boulder, Boulder, CO, USA. Email: \{firstname.lastname\}@colorado.edu.}%
\thanks{\textsuperscript{2}AJR is with the U.S. Army Combat Capabilities Development Command (DEVCOM) Army Research Laboratory, Aberdeen, MD, USA and with the Warfighter Effectiveness Research Center, United States Air Force Academy, Colorado Springs, CO, USA. Email: \{firstname.j.lastname.civ\}@army.mil.}%
}

\begin{document}

\maketitle
\thispagestyle{empty}
\pagestyle{empty}

\begin{abstract}

Effective collaboration between humans and AIs hinges on transparent communication and alignment of mental models. However, \textsl{explicit}, \textsl{verbal} communication is not always feasible. Under such circumstances, human-human teams often depend on \textsl{implicit}, \textsl{nonverbal} cues to glean important information about their teammates such as intent and expertise, thereby bolstering team alignment and adaptability. Among these implicit cues, two of the most salient and fundamental are a human's actions in the environment and their visual attention.
In this paper, we present a novel method to combine eye gaze data and behavioral data, and evaluate their respective predictive power for human proficiency,  trust, and  intent. We first collect a dataset of paired eye gaze and gameplay data in the fast-paced collaborative ``Overcooked" environment. We then train models on this dataset to compare how the predictive powers differ between gaze data, gameplay data, and their combination. We additionally compare our method to prior works that aggregate eye gaze data and demonstrate how these aggregation methods can substantially reduce the predictive ability of eye gaze. Our results indicate that, while eye gaze data and gameplay data excel in different situations, a model that integrates both types consistently outperforms all baselines. This work paves the way for developing intuitive and responsive agents that can efficiently adapt to new teammates. 


\end{abstract}

\section{INTRODUCTION}

With the continued advancement of artificial intelligence (AI) and robotics, it has become increasingly important to develop autonomous agents that can effectively collaborate with humans. One promising research direction focuses on endowing agents with a theory of mind \cite{machine_tom}, involving the development of mental models of teammates to improve adaptability \cite{mental_models}.
\textsl{``Explicit''} communication---which is direct, unambiguous, and oftentimes verbal---can help form such mental models  \cite{explicit_comm}.
However, in many real-world teaming scenarios, only \textsl{``implicit''} communication---which is indirect, suggestive, and often non-verbal---may be possible. This could be due to factors such as 
the need for rapid action execution
or high levels of ambient noise.
In these scenarios, autonomous agents must rely on implicit signals to understand their teammates.
Two implicit signals have been identified in the literature as promising options for these scenarios: 1) a teammate's behavior in the environment \cite{herding_pred}, which informs about intent and can anticipate future behavior and 2) a persons's visual attention, which provides fine-grained, immediate signals about their focus \cite{eyegaze_review}. 
While these signals have been leveraged independently to model and predict human behavior, few works have sought to combine them.
In this work, we hypothesize that by integrating these streams a more nuanced and complete model of a teammate can be learned. 
It is worth noting that acquiring high-fidelity data of complex cooperative tasks in sufficient quantities for deep learning models and comprehensive analysis is still a challenge in the field. This paper not only provides such a dataset collected through a large user study, but also provides a state-of-the-art (SotA) framework in the form of a causal transformer \cite{Radford2019LanguageMA} and analysis comparing different implicit signals to predict a human's: a) proficiency at a task; b) trust in an autonomous teammate; and c) future intents. 


\begin{figure}
\centering
\includegraphics[width=\columnwidth]{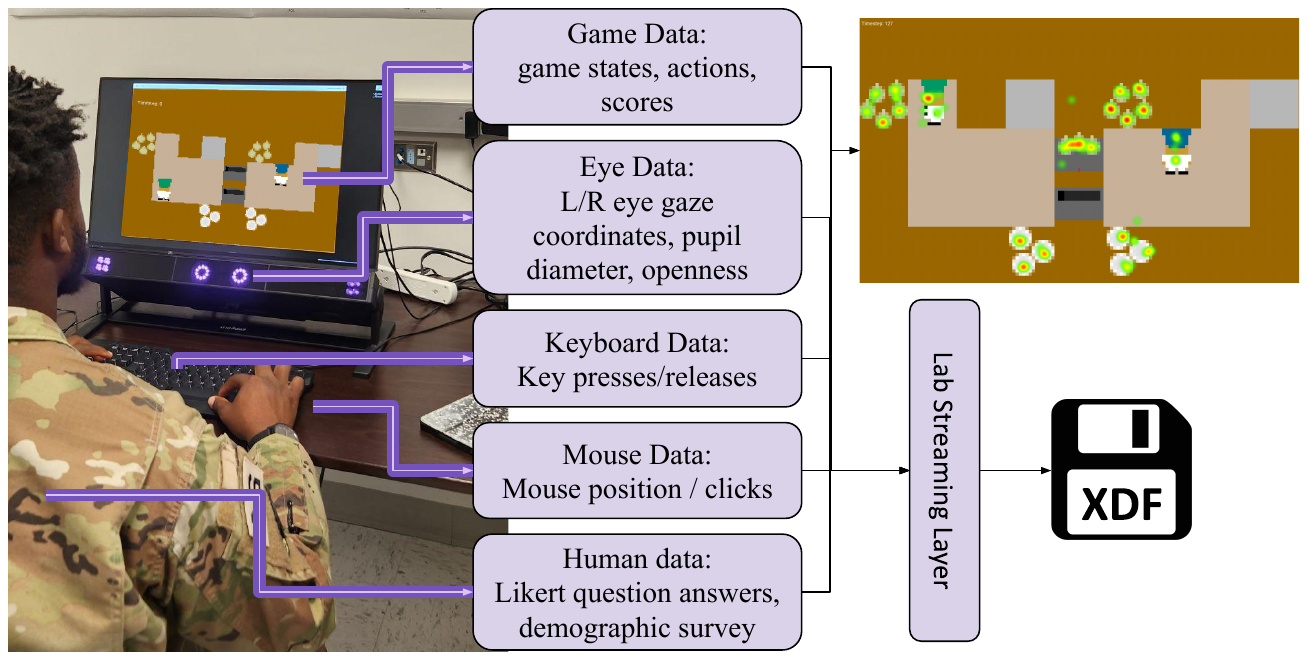} 
\caption{In this work, collect a large dataset of paired eye gaze and gameplay data in the collaborative game "Overcooked." Using this data, we train a causal transformer demonstrating state-of-the-art performance in its ability to predict a collaborator's task proficiency, trust in an autonomous teammate, and future intent.}  
\label{fig:setup}
\end{figure}


Prior work in human-robot interaction (HRI) and human-computer interaction (HCI) have demonstrated the predictive power of implicit signals like eye gaze \cite{trust, real_time, hand_eye} and behavioral data \cite{markov, human_info1, RGB}. Despite these advances, existing work still has limitations. First, most of these works infer only a single data point about their human teammates rather than build a comprehensive model of their behavior.
Second, they are often applied to non-representative, turn-based environments where a single action can span several seconds and the autonomous agent is limited compared to the human teammate.
Third, when eye gaze data is employed, it is often hand-crafted into a small set of features before being fed to predictive models, degrading a rich source of information for cluing the agent into the users' mental state. Finally, most of these papers do not publicly release their datasets, hindering replication, comparison, and further research. 

In contrast, this work not only leverages eye-tracking and behavioral data \textsl{in parallel} to accurately predict multiple latent human factors, but also performs a comprehensive analysis of these inputs
to determine the advantages of each data type.
We collect this data in the fast-paced collaborative ``Overcooked" environment (cf. see \cref{fig:setup} and \cite{oai}), which serves as an ideal testbed for human-AI teaming due to its capacity to efficiently gather large amounts of behavioral data in the form of intricate and coordinated gameplay at different levels of abstraction. 
We then leverage state-of-the-art deep learning models to predict multiple mental and behavioral factors including the human's intent (in the form of future attempted subtasks), their trust in the autonomous teammate, and their proficiency at the game.
Additionally, we compare several methods to aggregate and represent eye gaze data, finding that gaze data provides salient information faster than gameplay data, but that gameplay provides a stronger signal as the task progresses. Combining the two consistently matches or outperforms the individual signals.
Our results also show that gaze aggregation across the temporal dimension only minimally impacts results in our tasks, while the spatial aggregation method used in \cite{wachowiak} substantially worsens performance.


In conclusion, we present the following contributions:
1) a time-series model that can be conditioned on both human eye gaze and gameplay data for accurate predictions of behavioral intents, skill level and trust in the agent;
2) a thorough analysis comparing the predictive power of gameplay data, eye gaze data, and the combination of the two, providing practical insights the contexts in which each type of data is most effective; and,
3) a publicly released dataset of gameplay data paired with eye gaze data in a fast-paced collaborative environment. 
We believe these insights derived will enable AI agents to better model human teammates, allowing faster and more specific adaptation to improve the team fluency and capability.
By equipping agents with the ability to process implicit signals, we introduce new modes of understanding and expand the boundaries of human-agent interaction.

\section{RELATED WORK}
Implicit human
signals, such as 
EEG signals \cite{implicitEEG}, heart rate \cite{impliciteHeartRate}, 
recent actions \cite{markov},  body language cues \cite{face_body_sig}, and eye gaze \cite{eyegaze_review}, have been studied as a means to improve the human-machine interaction \cite{implicit_HCI}.

\paragraph*{Human behavior as a predictor}
Human behavior often contains informative action cues that hint at future intent.
For instance, reaching for a door handle suggests the intention to exit a room.
Notably, past sequences of behavior have been used to improve human-robot collaboration on assembly tasks \cite{col_ass, col_ass2}, anticipate a human's action in a herding game \cite{herding_pred}, enhance human performance in teleoperation tasks \cite{teleop}, and predict a decision making in search and rescue \cite{SaR}.
Other work has investigated how to predict an action based on an observed initial portion of it. Wearable devices have been employed to collect arm movement data and improve prediction in handover tasks \cite{handover}. Progress has also been made on predicting actions from RGB images and optical flow \cite{RGB} or RGB images alone \cite{rgb_only}, as well as on breaking down human movement into granular ``movemes'' to improve behavioral predictions \cite{moveme}.


\paragraph*{Eye gaze as a predictor}
Eye gaze stands out as a salient signal, providing rich insights into a person's attention, information processing, and social interactions, enhancing human teaming 
\cite{humanhuman_eg}. 
It has been used to anticipate intent in a robotic manipulation tasks \cite{eyegaze1}, as a substitute for wake-words for smart-speakers \cite{tama}, predict train routes in a turn-taking train board game \cite{tickettoride}, and detect errors in robot behavior \cite{eg_error}.
Interestingly, the predictive power of gaze has been also demonstrated on the other end of the human-robot dyad: a robot equipped with a human-like binocular system and corresponding gaze controller \cite{roncone2016cartesian} improves the human's ability to predict the robot's intent \cite{boucher2012reach}.

Our work shares similarities to \cite{candon} and \cite{wachowiak}. Both works use implicit signals, including eye gaze, to predict information about humans in a fast-paced collaborative environment. \cite{candon} explores the use of gaze features, game data, survey data, and demographics to predict users' preferences between early game assistance and late game assistance. However, unlike our study, this work does not compare the use of gameplay data on its own to eye gaze data alone.
\cite{wachowiak} uses eye gaze data to predict periods of human confusion in the same environment we employ, however they do not consider the use of gameplay in any form.
Notably, both of these works aggregate gaze data over both time and space. Our research differs by 1) thoroughly comparing gaze data to gameplay data, 2) examining the effects of aggregating gaze data in multiple different ways, and 3) exploring the predictive power of implicit communication across the three differentdimensions of trust, proficiency, and intent.

\section{METHOD}
\subsection{Data Collection}
\subsubsection{Environment}
Due to its highly flexibly nature and its ability to capture a wide-range of human-agent teaming behaviors, we focus our work on the collaborative cooking game ``Overcooked'' \cite{oai}. ``Overcooked" requires a team composed of a human and an AI-controlled chefs to cook and serve as many soups as possible within a set time limit. To achieve this, players must execute a series of tasks ranging from collecting onions to placing them in a pot and serving the finished soups. Successful service rewards the team with 20 points.
At each timestep, each player can choose one of the following base actions: \textit{up}, \textit{down}, \textit{left}, \textit{right}, \textit{interact} with an object (to pick up, place, or serve items), or \textit{stay} in place. ``Overcooked" requires players to coordinate both on high-level strategic decision and on their underlying movements. At the strategic level, players should aim to minimize redundancy and inefficiency—for instance, avoiding the situation where both players retrieve a dish when only one soup is being prepared. On the movement level, careful navigation is essential to prevent collisions between players. This combination of strategic planning and movement precision makes ``Overcooked" an especially suitable platform for studying human-agent collaboration.



\begin{figure}
\centering
\subfloat[\centering Asymmetric Advantages]{{\includegraphics[height=2cm]{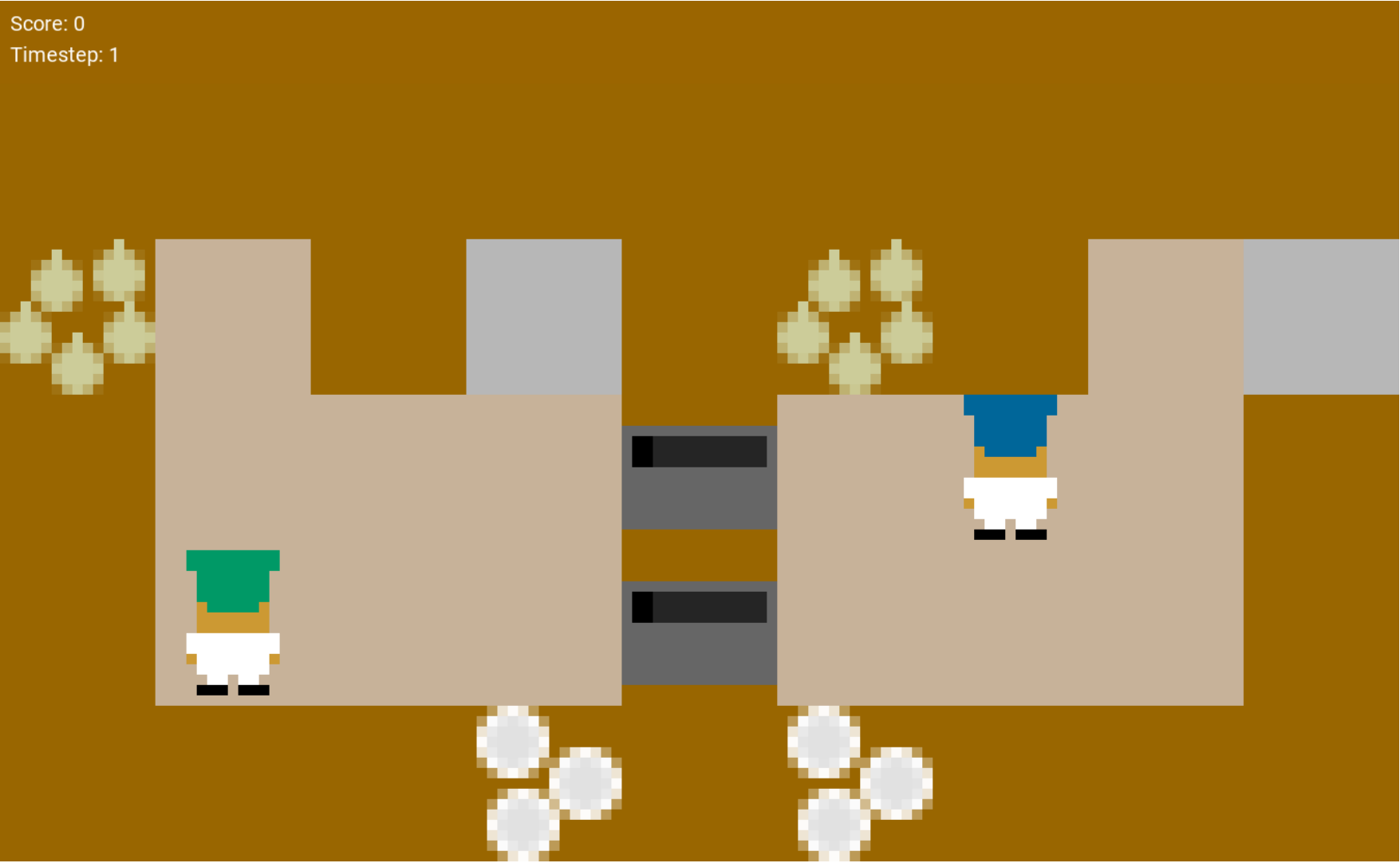} }}
\subfloat[\centering Coord. Ring]{{\includegraphics[height=2cm]{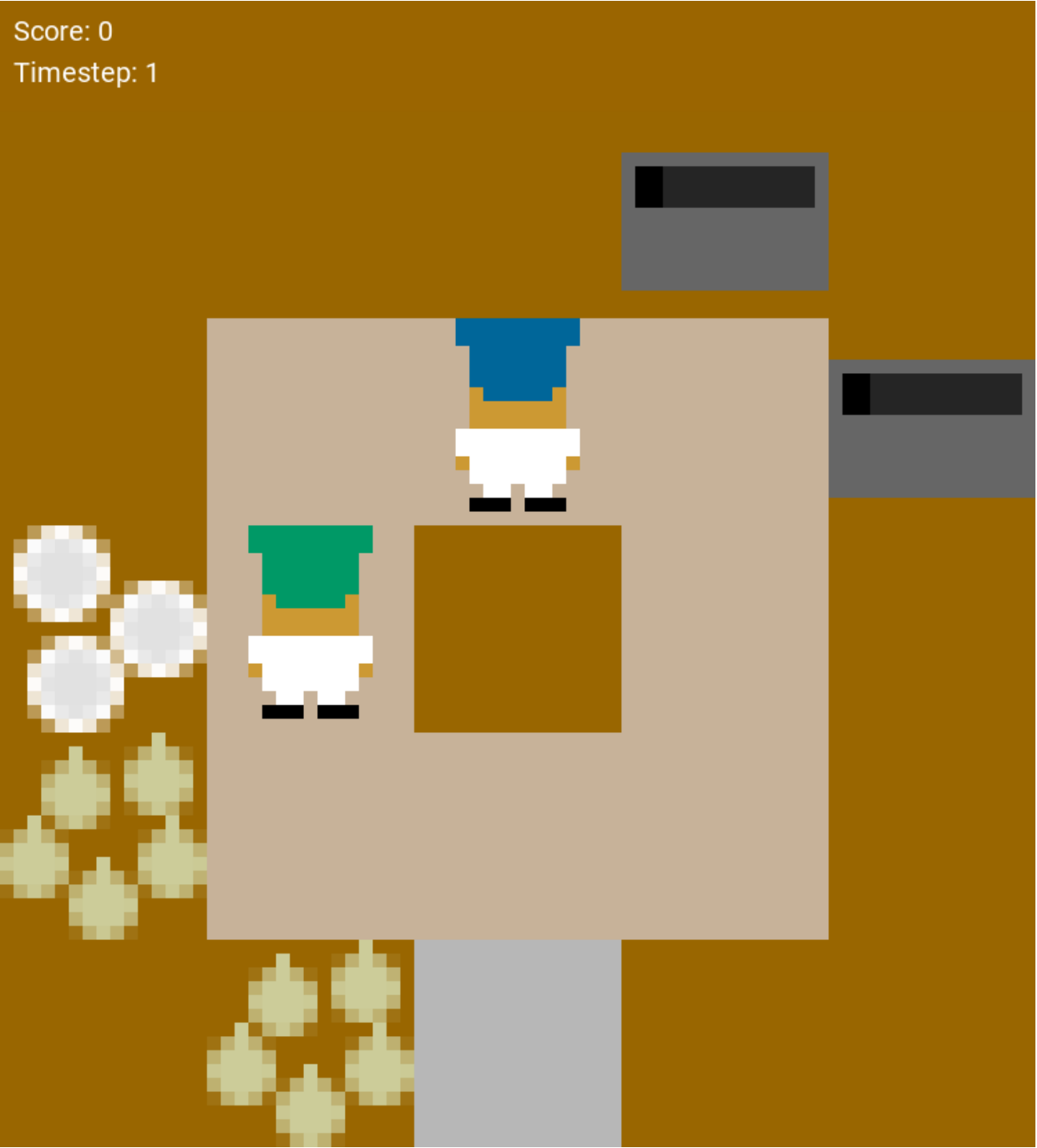} }}
\subfloat[\centering Counter Circuit]{{\includegraphics[height=2cm]{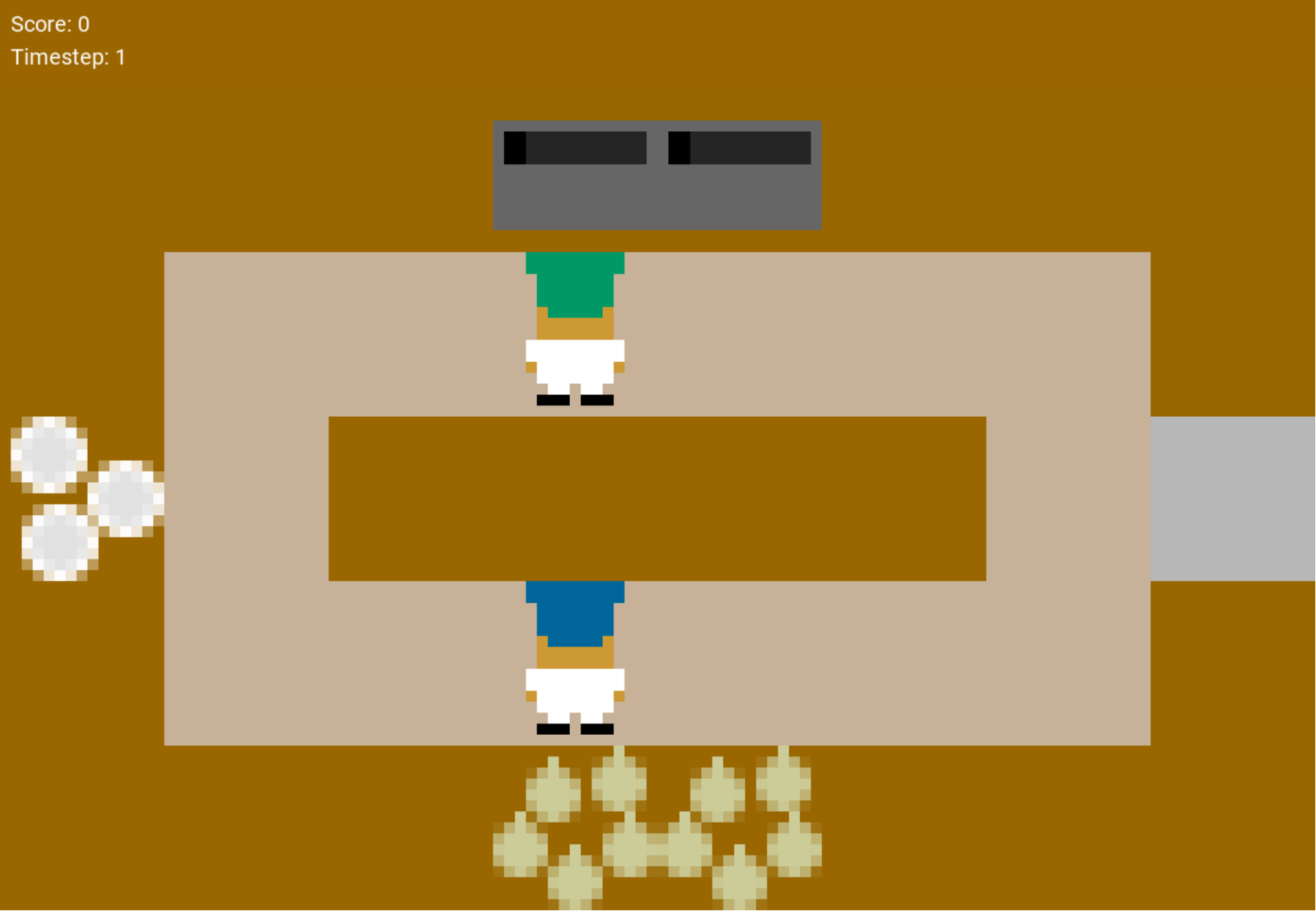} }}
\caption{The three ``Overcooked" layouts used. From \cite{oai}.}
\label{fig:3envs}
\end{figure}

\cref{fig:3envs} shows the three specific game layouts we use to gather data: 1) Coordination Ring, which requires agents to focus heavily on their movement to avoid collisions 2) Asymmetric Advantages where agents are fully separated and so cannot collide, but must instead focus on aligning their high-level strategy, and 3) Counter Circuit that requires both movement and strategic alignment. Following previous work \cite{haha}, we ran the experiments for $400$ in-game timesteps at $5$ FPS, which equates to $80$ seconds of gameplay.



\subsubsection{AI Agents}
To capture a thorough and wide range of human behaviors, we collected data using three different agents of varying ability. The first agent is a random agent, which randomly selects one of the six base actions. This represents a very low level of play and is intended to create situations where the human may be confused about its teammate leading to low trust. The second agent is a self-play (SP) agent that is trained using reinforcement learning (RL)---specifically proximal policy optimization (PPO) \cite{PPO}---and, as the name suggest, is trained being teamed with itself. This agent can be quite good at the game if the human adapts to its play-style, but its training regime causes it to be very rigid in its behaviors. This agent is aimed to create trials where the human can have moderate trust in their teammate, but must still pay attention to the agent's behavior to avoid frequent collisions and a lower final score. Lastly, we use a SotA HAHA agent \cite{haha} that has been shown to be a significantly more performant, trusted, and understandable teammate. This agent was included to elicit situations where the humans have a high-level of trust in their teammate.

\begin{figure*}
    \centering
    \includegraphics[width=500pt]{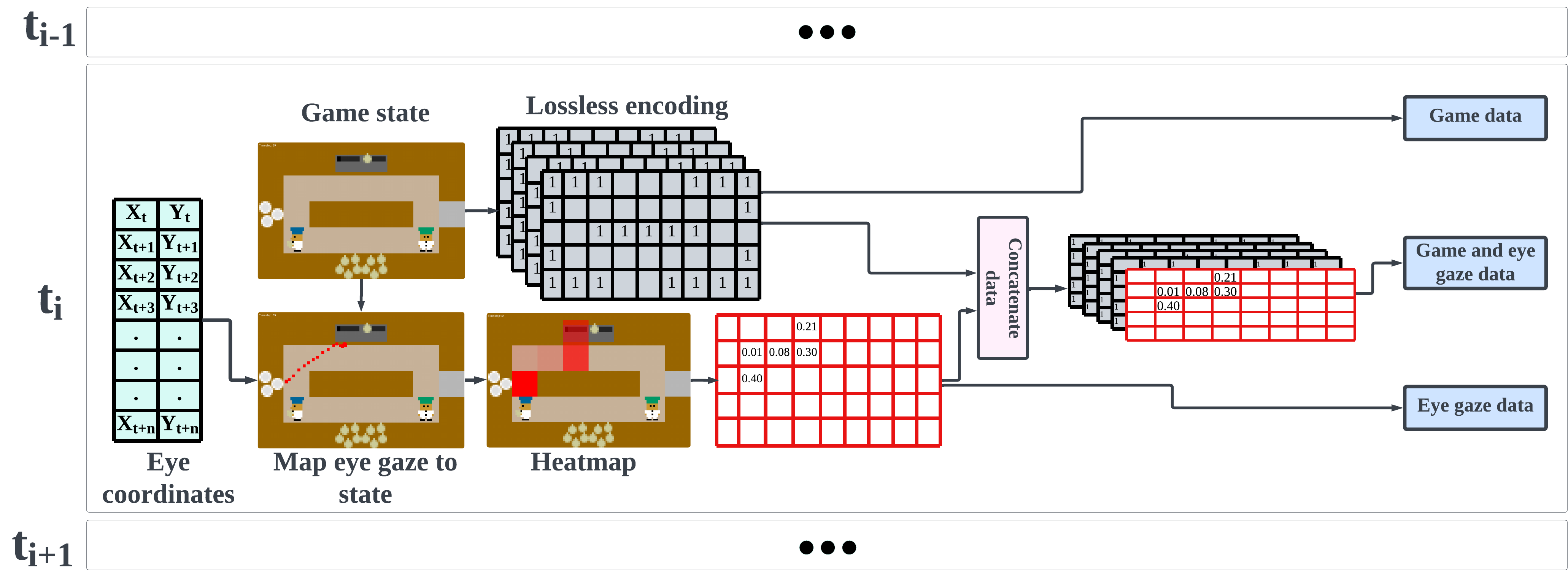}
    \caption{An overview of the processing method to create representations of eye gaze data, gameplay data, and enable a combination of the two for a single timestep. The representations are designed to be easily fed into modern neural networks.}
    \label{fig:arch}
\end{figure*}

\subsubsection{Trial Design}
The primary objective in the dataset creation was to collect a wide range of human behaviors while performing a collaborative task from which we could analyze and compare the predictive ability of gameplay data and human eye gaze data. To this end, we ran an IRB-approved user study where we recorded participants playing the collaborative cooking game Overcooked. After completing consent forms, participants were required to fill out a demographic survey, read instructions about the game, and complete a short tutorial that required them to serve a completed soup before moving on. Each participant then played 18 rounds of Overcooked, with each round being played with one of three different agents on one of three different layouts. This led to each participant playing the nine different layout-agent combinations twice during the full duration of the trial. In the next section, we provide a more in depth description of the Overcooked environment and the specific layouts used, the set of different agents used, the number and recruitment methods for participants, the data collected during each trial, and how the data was processed.

\subsubsection{Participants} 
In total, 83 participants were recruited across both the United States Air Force Academy (USAFA) and the University of Colorado (CU) Boulder using newsletter announcements and an online recruiting software. Nine participants were removed due to either technical difficulties with the system or poor eye tracking data quality ($>40\%$ of eye tracking data was missing on at least one trial), leaving $74$ participants in the dataset, for a total of $1332$ total rounds of play or $29.6$ hours of recorded play time. The age of participants ranged from 18 to 52 with an average age of 21.43. 33 participants identified as male, 39 as female, 1 non-binary, and 1 preferred not to disclose. When asked about their previous experience with Overcooked on a scale of 1 to 7 (1 being no experience), participants reported an average of 1.45, indicating that the majority of our participants had no or limited experience with the game. As our primary objective is to test predictive ability with unseen humans, we randomly select $59$ participants for our training set, $5$ participants for our validation set, and $10$ participants for our test set. All participants were required to have normal vision ($20/40$ or better) without contact lenses to ensure that the eye capture system would be effective. Prior to participating, volunteers signed an informed consent document approved by the IRB at the Army Research Laboratory (ARL 23-079) in accordance with the Declaration of Helsinki. 

\subsubsection{User Study}
Participant were required to complete an online demographic survey prior to their in person session.

Upon arrival, participants signed a consent form, and were then positioned around $70$cm away from a display and attached eye-tracking device (Tobii ProSpectrum), at which point a five-point calibration was executed using the Tobii Eye Tracker Manager (2.6.0) \cref{fig:setup}. Subsequent to calibration, the accuracy of eye tracking was confirmed via real-time gaze tracking, with mandatory recalibration for any validation point discrepancies exceeding $1.5^{\circ}$. 

Following this, a Lab Streaming Layer (LSL) stream was started that broadcasted the eye gaze data (including, but not limited to, the right and left eye $x$ and $y$ coordinates relative to the display, as well as pupil dilation collected at $300Hz$), the game data (including, but not limited to, the game states, team actions, the reward, and instance of collisions), and the keyboard data and mouse data. All data was recorded in xdf files. See \cref{fig:setup} for depiction of the setup. Between each round, we additionally collected the participants answers on five statements adapted from \cite{hofman} using a 7-point Likert scale \cite{likert} ranging from strong disagreement to strong agreement. These statements pertained to team fluency, perceived role significance, trust in the agent, understanding of agent actions, and the agent's cooperativeness. 

\subsection{Data Processing}

To enable information to be readily fed into neural networks, we first clean and process collected data. We use \cite{oai}'s lossless state encoding function to encode the game states into a grid representation of shape height x width x $27$, where each of the $27$ channels contains information about different in-game objects or players. For the eye gaze data, we first average the $x$ and $y$ pixel coordinates of the two eyes. If the gaze data for a single eye is null for a given sample, we use the data from the only valid single eye data. If the gaze data is null for both eyes for a given sample, we exclude that sample. We then map the pixel coordinates to the corresponding tile in the game's grid environment. During this process, we filter out all eye gaze samples where the participant is not looking within the boundaries of the game environment. Since the eye gaze data is sampled at roughly 300Hz compared to the 5Hz (or FPS) of the gameplay data, we have approximately $60$ eye gaze data points per gameplay timestep. To enable the combination of gameplay data and eye gaze data, we create eye gaze heatmaps of the same shape as the underlying game grid, and populate the grid with the ratio of gaze samples that fall within the boundaries of each tile. A visual representation of our method is shown in \cref{fig:arch}. To compare our method to the method used in \cite{wachowiak}, we additionally map each eye gaze sample to a game grid tile and classify the sample as the human looking at their own agent, looking at their teammate, or looking at the environment. For each game timestep, we calculate the ratio of samples in each of these three bins.

After processing, we can readily use five input representations for a given number of timesteps. 1) a lossless game state encoding per game timestep (Game Data), 2) an eye gaze heatmap per game timestep (Eye Gaze Data), 3) a combined state encoding and the eye gaze heatmap per timestep (Game Data + Eye Gaze Data), 4) the average heatmap across all twenty timesteps (Collapsed Eye Gaze), and 5) the average ratio of eye gaze samples that map to the human's agent, teammate, environment across all timesteps (Gaze Object).

We use three different labels from our dataset. The first are the humans levels of agreement on the likert question: "I trusted the agent to do the right thing:". This ranges from $0$ (strongly disagree) to $6$ (strongly agree) with $3$ being neutral. Second, for each agent-layout pair, we bin all scores in tertiles and label rounds by the tertiles they scored in. Scores in the bottom tertile would be in bin $0$ (beginners), the middle tertile in bin $1$ (intermediates), and the top tertile in bin $2$ (experts). We use these score tertiles as a proxy for human proficiency. Third, we calculate when the human player completes one of the eleven different subtasks by identifying each time they perform an \textit{interact} action and inspecting the change in state. We then back label each timestep since the previous subtask completion with the completed subtask. We use these subtask labels to predict a human's future intents. We note here that for trust and proficiency, there is a single label for a full $400$ timestep round. For intent, there are many subtask labels in a single round, and the duration of a subtask label is highly variable and dependent on which task is being performed, the current layout, and the proficiency of the human. 

As this data is collected from a human study and not hand curated, class distributions are not perfectly balanced. Reporting accuracy in this case can over state the performance. Due to this, we use an F1 score as our primary metric, as F1 scores incorporate both precision and recall in its final output. We additionally include a baseline model that always predicts the majority class in all our results.

\begin{figure*}[h]
\centering
\begin{tabular}{ccc}
  \subfloat[Proficiency Prediction]{\includegraphics[width=0.30\textwidth]{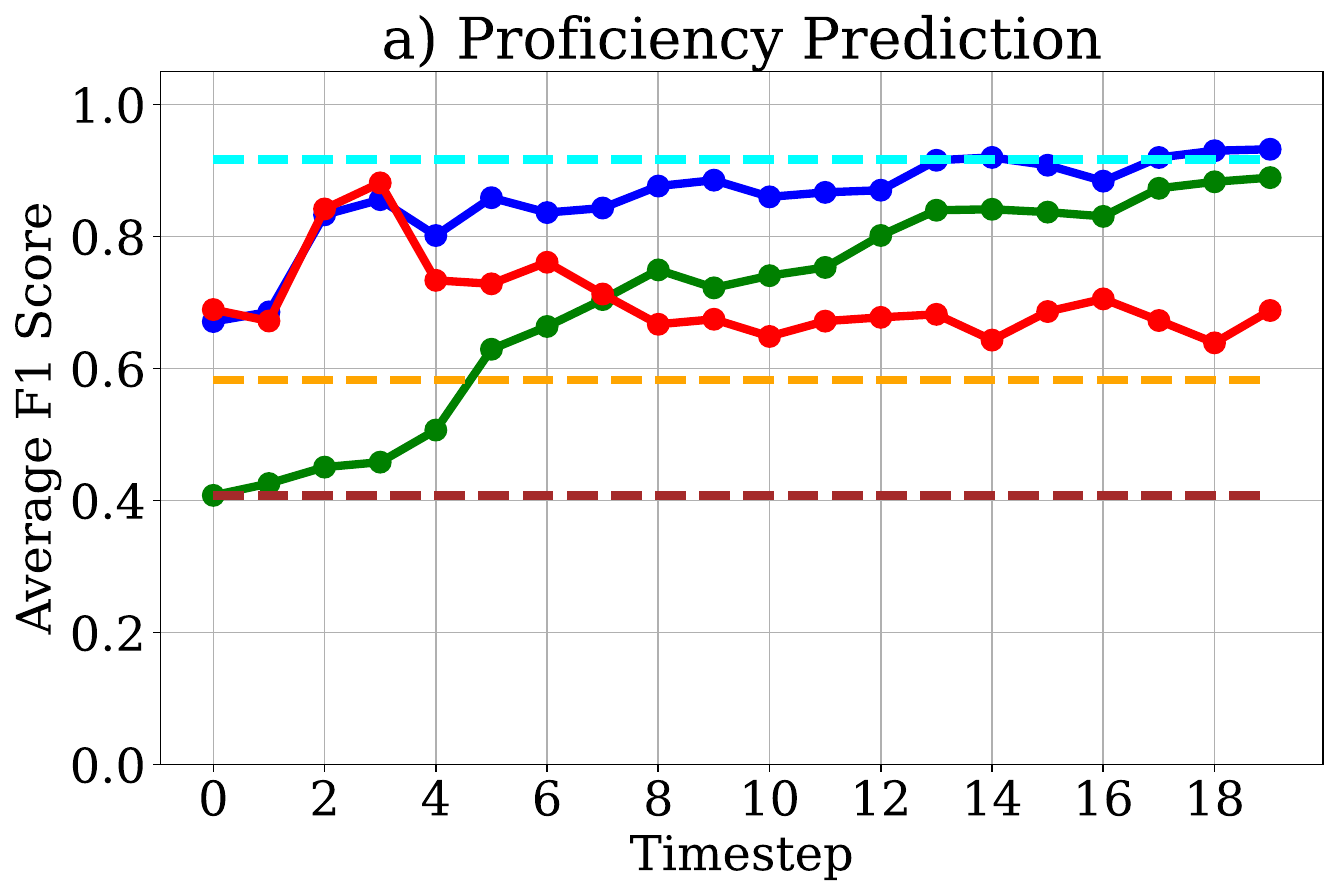}} &
  \subfloat[Trust Prediction]{\includegraphics[width=0.30\textwidth]{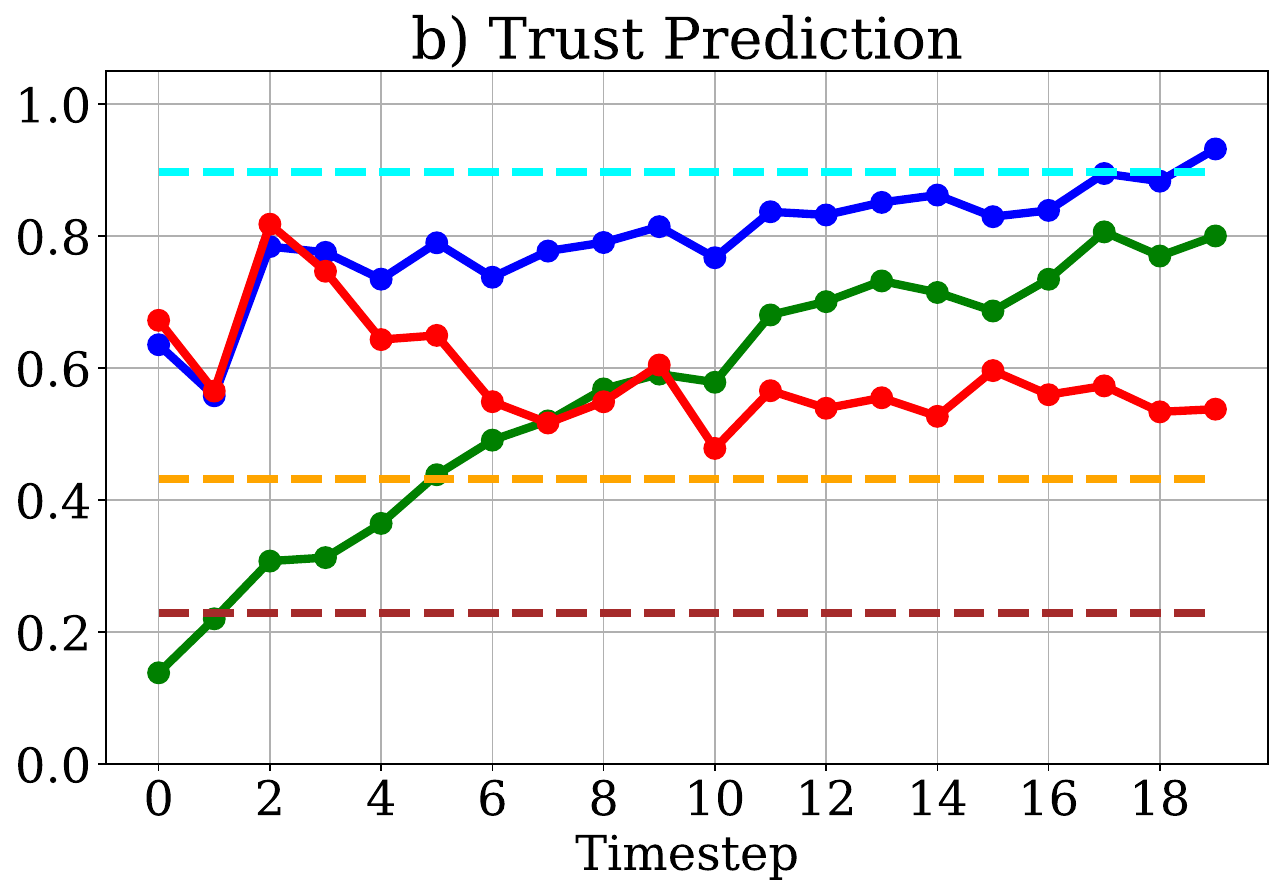}} &
  \subfloat[Intention Prediction]{\includegraphics[width=0.30\textwidth]{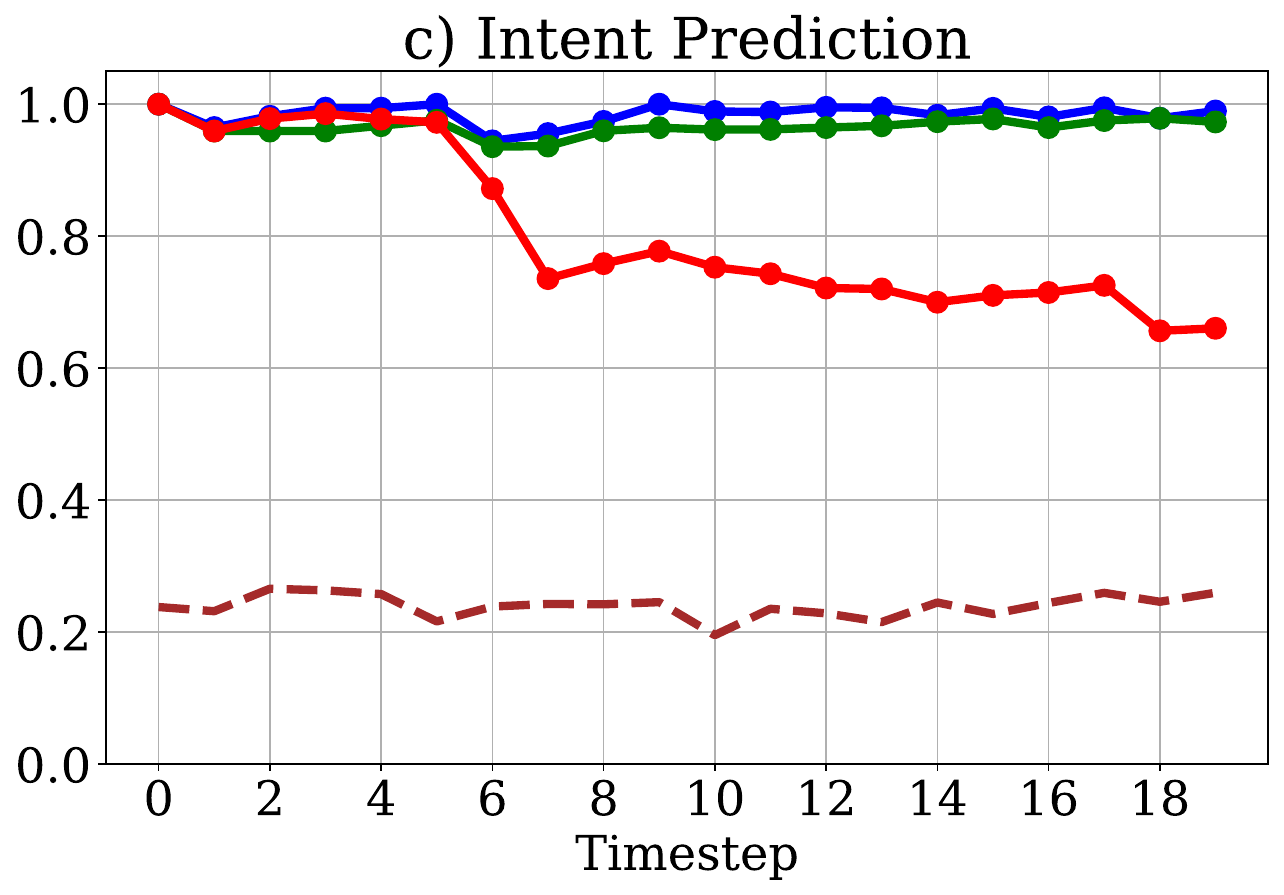}} \\
  \subfloat[Proficiency Cumulative F1]{\includegraphics[width=0.30\textwidth]{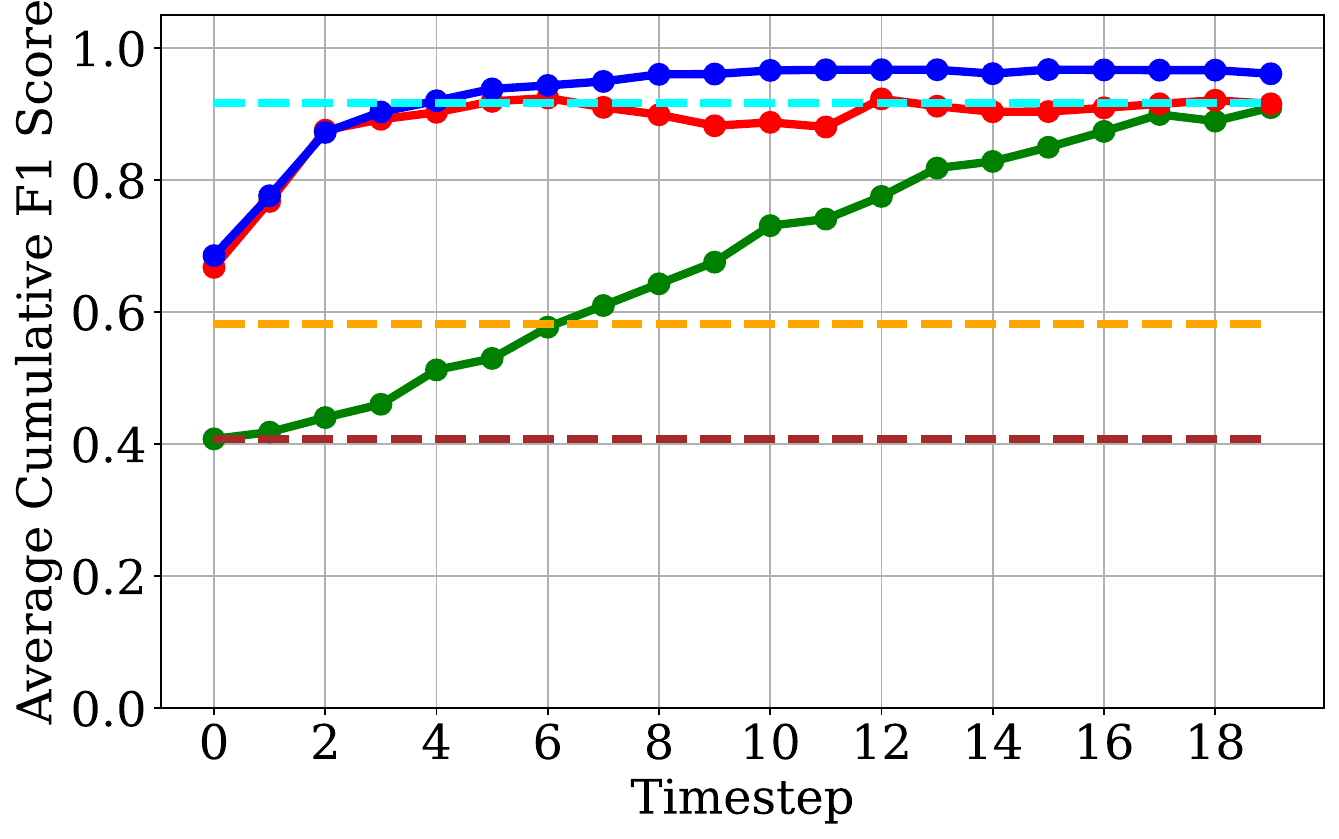}} &
  \subfloat[Trust Cumulative F1]{\includegraphics[width=0.30\textwidth]{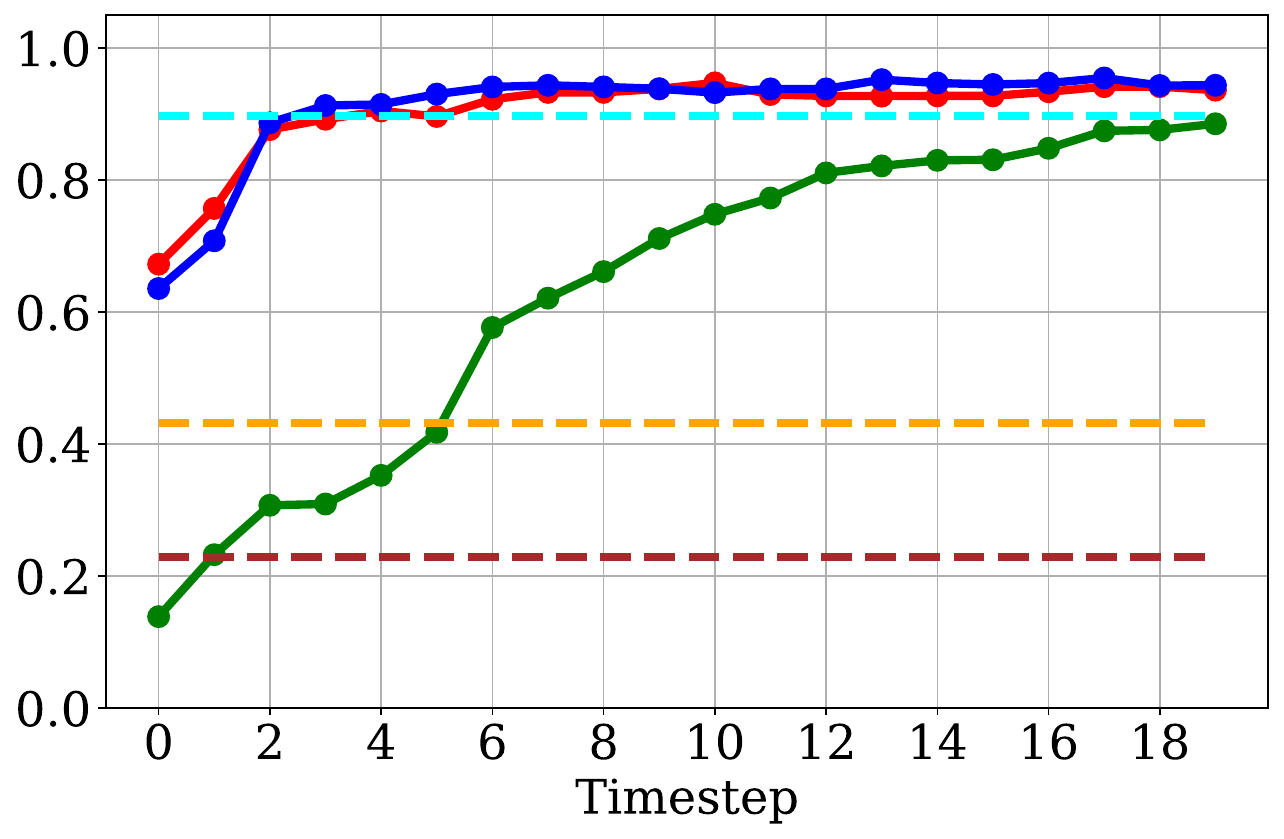}} &
  \subfloat[Legend]{\includegraphics[width=0.30\textwidth]{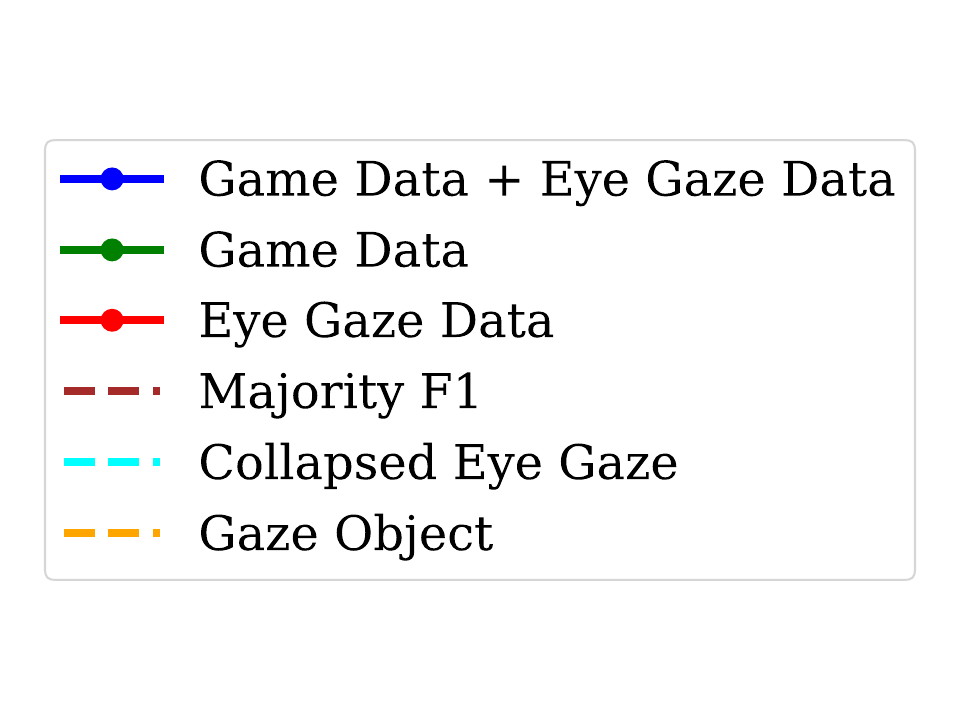}}
\end{tabular}
\caption{F1 scores over time for different implicit human signals predicting human proficiency, trust, and future intents starting at timestep $\mathbf{0}$ of each trial. The top row of graphs shows the per-timestep prediction outputted by our transformer model that can handle time-series data. The bottom row shows the cumulative prediction of all past timesteps. Dotted lines represent methods that aggregate over time and use the full $20$ second window for their prediction.}
\label{fig:main_results}
\end{figure*}

\subsection{Models}
We train two types of models to predict our labels from the input data. For the input data types that retain time-series information---game data, eye gaze data, game data + eye gaze data---we first flatten the timestep representations, as in \cref{fig:arch}. We then apply a linear layer to encode them into a token embedding size and pass the first 20 timesteps through a transformer model \cite{transformer}. To capture the temporal dependencies in the data, we employ a causal transformer architecture \cite{Radford2019LanguageMA}. Specifically, we generate a causal attention mask that ensures each output token can only attend to the previous tokens in the sequence. This masking mechanism is crucial for preventing information leakage from future timesteps and enabling the model to learn meaningful temporal patterns. To prevent overfitting, techniques like dropout and layer normalization are applied in positional encodings and transformer layers. Each output token is fed into a linear layer to get the appropriate number of logits for the task at hand. We use a cross-entropy loss between the logits and ground truth labels at every timesteps and the RAdam optimizer \cite{Radam} to train the model. We use the same architecture parameters as the base model in \cite{transformer}. We perform a grid search on learning rate: $lr \in \{1e-5, 3e-5, 1e-4\}$, batch size: $bs \in \{32, 64, 128\}$, warmup steps: $ws \in \{500, 1000, 2000\}$ and found $lr=3e-5$, $bs=128$, and $ws=2000$ provided the best results.

For the two representations that aggregate over timesteps---collapsed eye gaze and gaze object---we average their representations over all 20 timesteps and then feed the aggregated input into a three layer multi-layered perceptron with 128 hidden units. We use the same loss function and optimizer. We perform the same grid search excluding warmup steps which are transformer specific and found $lr=1e-4$ and $bs=128$ provided the best results.

\subsection{Data Release}
An anonymized version of the collected data and the code used to process it can be found online\footnote{\href{https://hiro-group.ronc.one/overcooked-eye-gaze-dataset}{\texttt{https://hiro-group.ronc.one/overcooked-eye-gaze\-dataset}} hosts the dataset. \href{https://github.com/HIRO-group/HAHA/tree/EyeGaze}{\texttt{https://github.com/HIRO-group\\/HAHA/tree/EyeGaze}} hosts the code used to process this data.}.  The dataset contains XDF files that include all eye gaze data at ~300Hz and all gameplay data at ~5Hz. Additionally, they include keyboard and mouse data that were not utilized in our analysis. In addition to the XDF files, the datset contains the results of the likert scale questions, which can be mapped to the XDF files using anonymized user and trial ids. 

\section{EXPERIMENTAL DESIGN}
With the collected data, we set out to answer the following three research questions.  
\textbf{RQ1: How does the predictive power of eye gaze data compare to the predictive power of gameplay data and to the combination of both?}
Core to our contributions is a thorough analysis of the predictive power of gaze data compared to gameplay data. To this end, we train a causal model per agent-layout combination on the first $20$ timesteps of each round for each of our three prediction labels: trust, proficiency, and next subtask to be completed.

\textbf{RQ2: How does aggregating eye gaze data along spatial and temporal dimensions effect its predictive power?}
Recent work has often aggregated eye gaze data across different dimensions to simplify the input space \cite{wachowiak,candon}. This immediately poses the question of if and by how much these simplifying aggregation techniques are impacting the predictive power of eye gaze data. To test this, we compare the predictive power when using the full time series eye gaze data to two lossy methods. In the first method, we average the heatmap across timesteps, which collapses the temporal dimension of the data and that we name \textit{collapsed eye gaze}. In the second,  inspired by the approach used in \cite{wachowiak}, we collapses the spatial dimension and only looks at the ratios of eye focus on the user themselves, the teammate, and the environment. We name this method \textit{gaze objects}.

\textbf{RQ3: Does the predictive power of eye gaze and gameplay data differ between the start of a new task and during continuous execution?}
Lastly, we hypothesize that a human's work flow may change between the start of a new task and when they have been performing the same task for a while. If true, we expect to see a difference in game play and gaze data patterns. To examine this, we compare the predictive power of eye gaze and gameplay data on when focusing on the first $20$ timesteps of gameplay compared to focusing on timesteps $200$ to $220$.

\section{RESULTS}

\textbf{RQ1: Comparing eye gaze data to gameplay data.} \cref{fig:main_results} depicts the predictive power of eye gaze data, gameplay data, and their combination across multiple human mental and behavioral factors. We first focus on the intent, or ``next subtask'' prediction, shown in \cref{fig:main_results} c). As this particular analysis only considers the inital $20$ time steps of the game, almost all participants will retrieve an onion as their first subtask, leading to a very high f1 score early on. However, whereas the models that use game data and the combination of game and gaze data maintain a high predictive ability, using ungrounded gaze data on its own leads to a drop in performance at later stages. This is likely because the gaze data only provides information on \textit{where} the human is looking, but without the game data, there is no information on \textit{what} the human is looking at. While the model can memorize the location of fixed objects in the environment to provide a better than random prediction, there is no way to know where either agent is situated or to model the dynamic changes to the environment. In a situation where the human looks at the location where a pot is, for example, from ungrounded eye data alone it would be difficult to ascertain whether the human is going to drop an onion into the pot, retrieve a completed soup from the pot, or perhaps check if their teammate is performing either of these actions.

We next focus on the proficiency and trust predictions, shown in \cref{fig:main_results} a) and b). For these, since the labels are identical for an entire round, we provide two versions of the graph. The top row of graphs show the individual predictions at each separate timestep, whereas the bottom row of graphs averages all the probabilities up to and including the current timestep. These results show a clear trend where the predictive power of gameplay data starts near majority class prediction, and consistently trends upwards. This is expected because the game's initial state is the same in all trials, and the model can refine its prediction as trajectories deviate toward or away from optimal paths. Notably, game data achieves a high performance within $20$ timesteps.

Eye gaze data alone provides a strong predictive signal very early on, spiking around timestep $3$ and $4$ in both the trust and proficiency predictions. A qualitative analysis of the eye gaze and behavior at the start of the game showed a trend where the participants would first look at their teammate, then switch their focus to their agent before performing their first productive action. For more experienced players, this shift would occur around this threshold of timestep $3-4$, whereas for less experienced players, it would occur later, usually between timesteps $7$ and $15$, aligning with the results we see here. Similarly to the subtask prediction, as the players deviate away from the start state, the gaze data lacks necessary game information and the models start to lose some of its predictive ability. However, as seen in the bottom row, this can be significantly mitigated by using the cumulative probabilities of all predictions. 

Lastly, using a combination of eye gaze data and gameplay data provides the best of both modalities, requiring little data to get a good performance, and continuing to improve with more data. Unlike the ungrounded gaze only models, the gaze here can be attributed to objects in the environment, and we see no drop in performance. In all cases, using both modalities provides the best or tied for best performance.

\begin{figure*}[ht!]
\centering
\begin{tabular}{ccc}
  \subfloat{\includegraphics[width=0.30\textwidth]{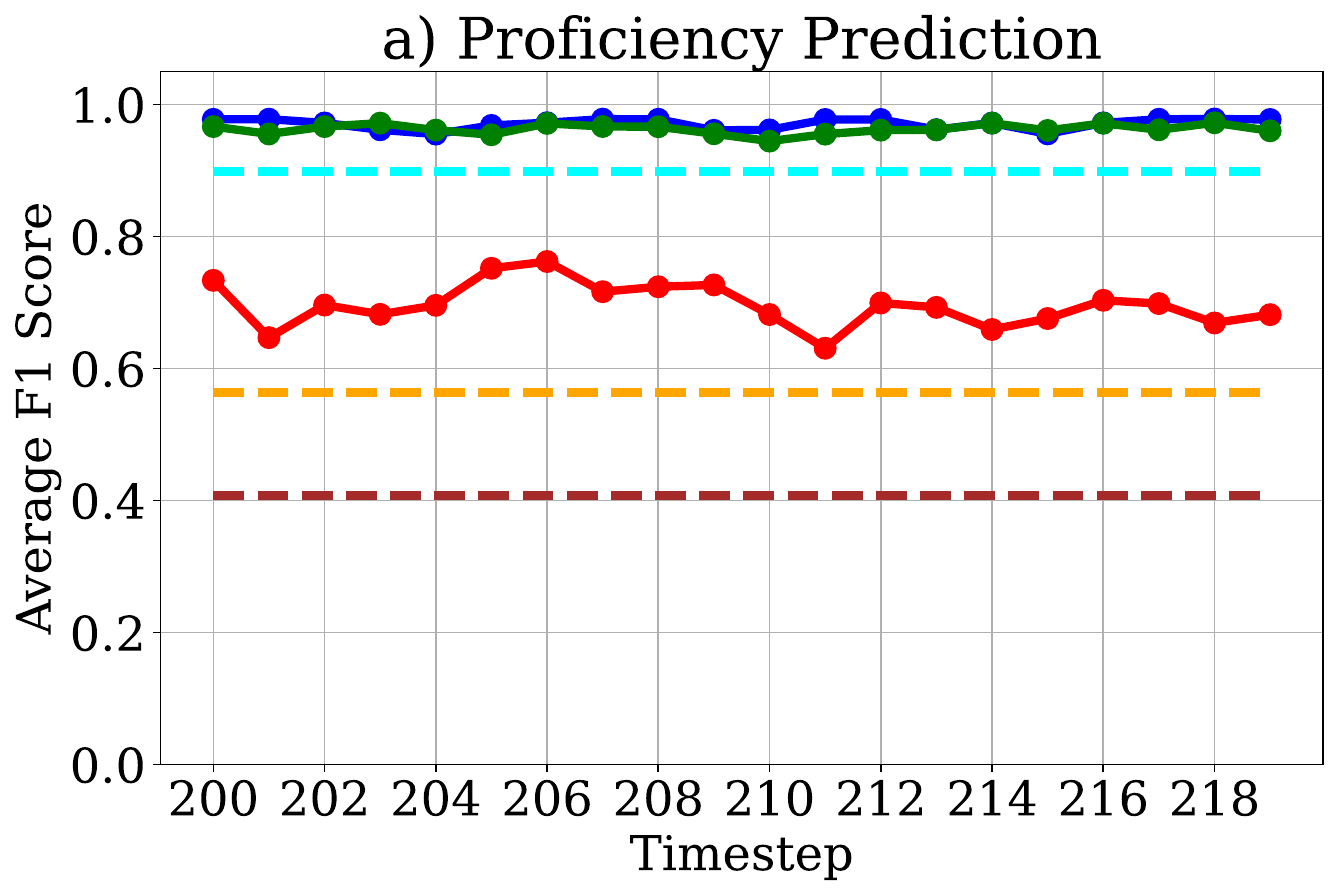}} &
  \subfloat{\includegraphics[width=0.30\textwidth]{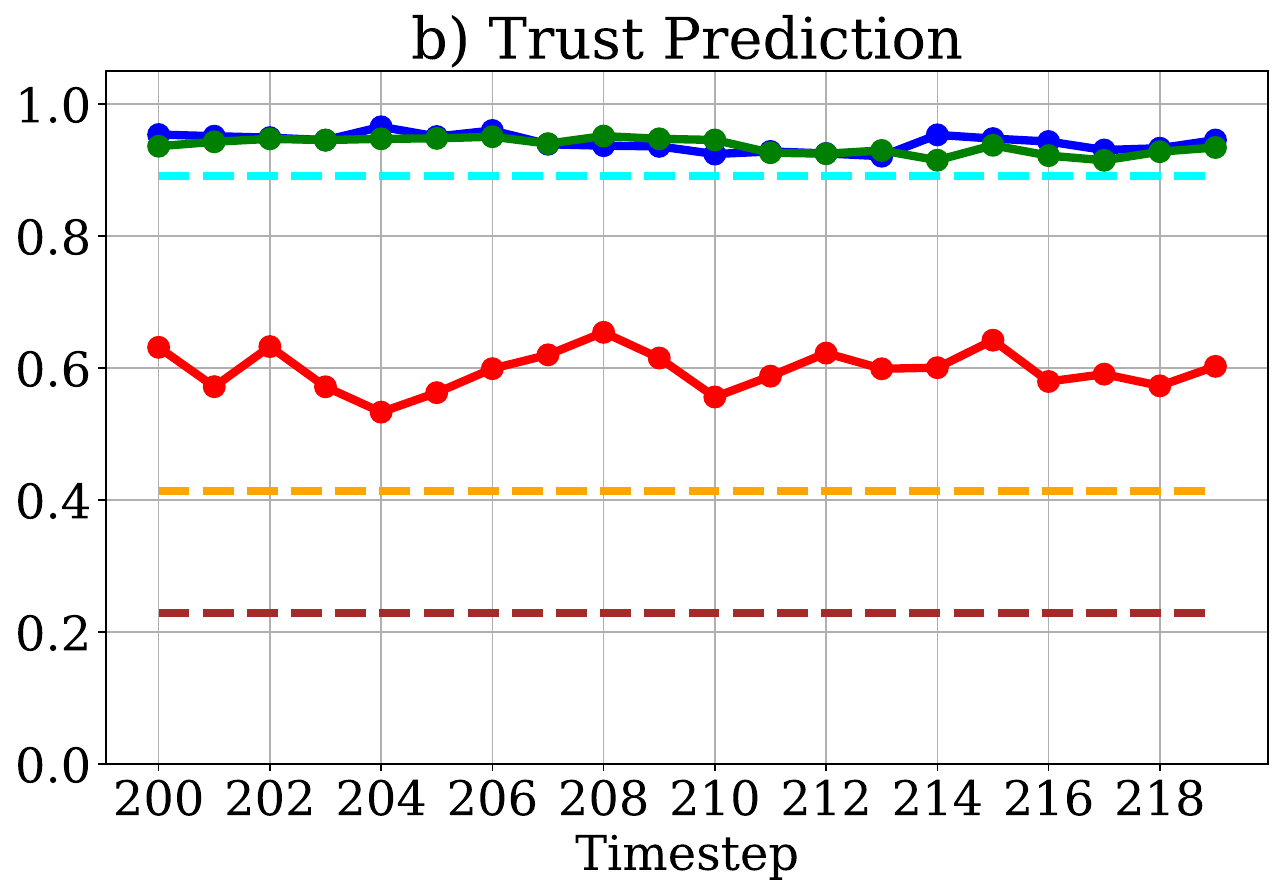}} &
  \subfloat{\includegraphics[width=0.30\textwidth]{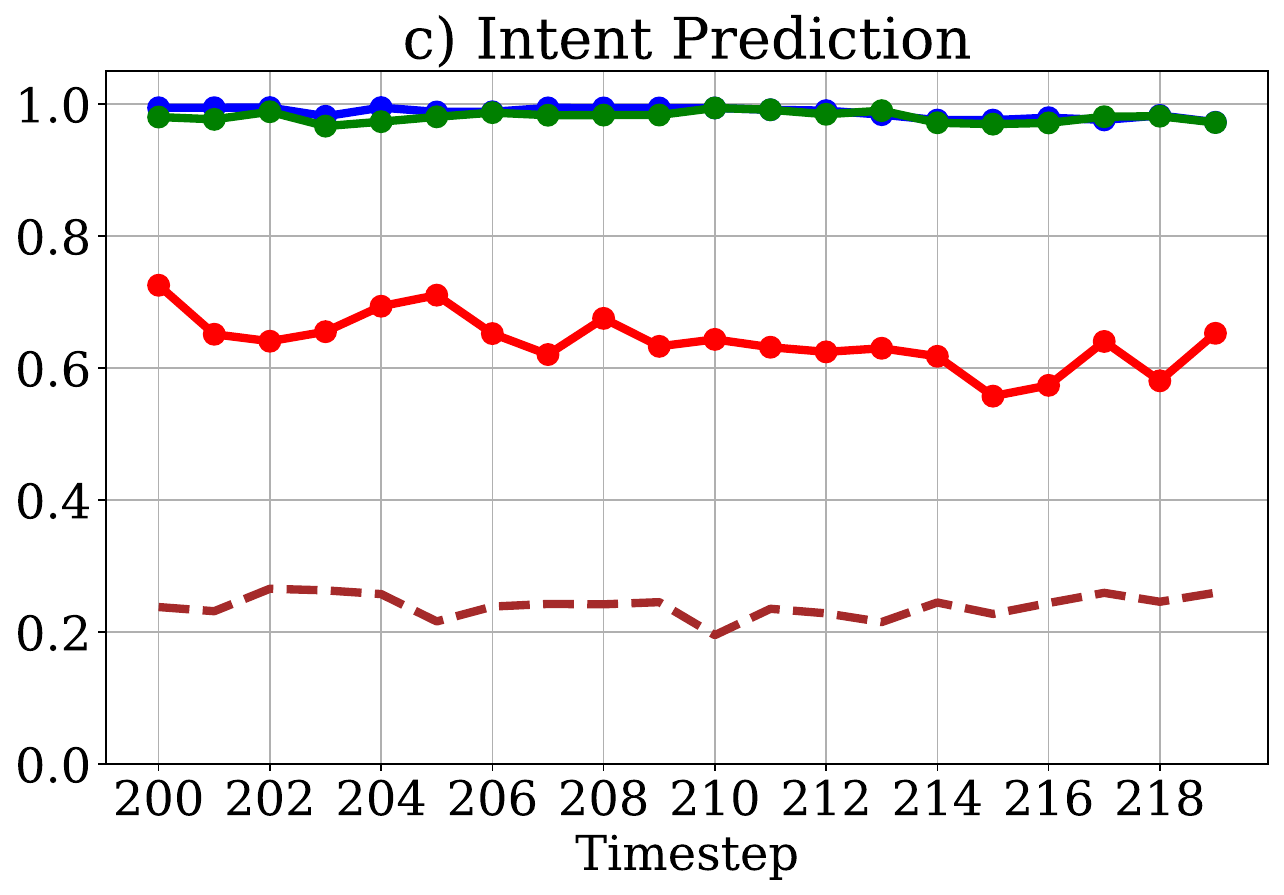}} \\
  \subfloat{\includegraphics[width=0.30\textwidth]{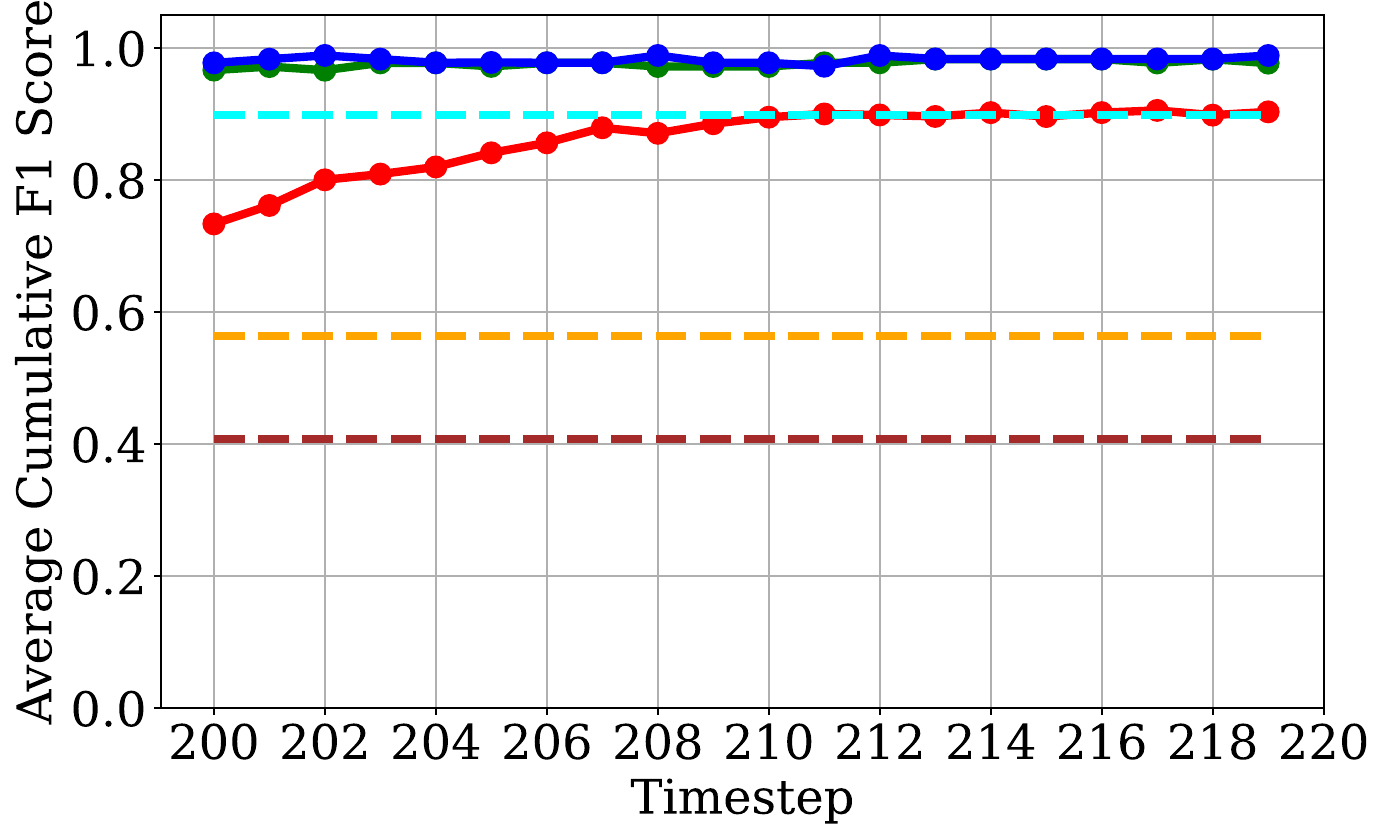}} &
  \subfloat{\includegraphics[width=0.30\textwidth]{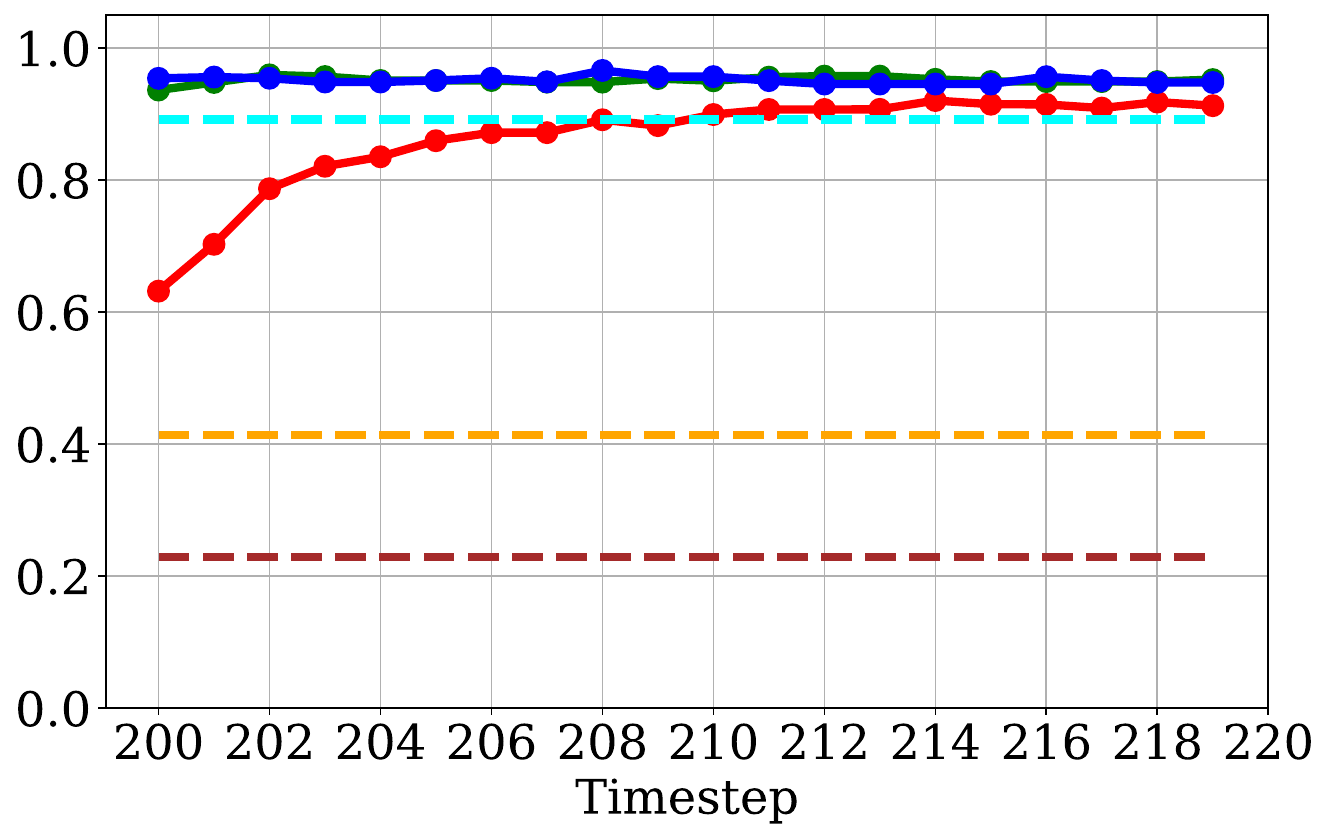}} &
  \subfloat{\includegraphics[width=0.30\textwidth]{figs/legend_svg-tex.pdf}}
\end{tabular}
\caption{F1 scores starting at timestep $\mathbf{200}$. Refer to \cref{fig:main_results} for a full description of the figure}
\label{fig:results_200}
\end{figure*}

\textbf{RQ2: The effect of gaze data representation.} We next investigate different methods to represent eye gaze data. Specifically, we compare the full time series representation utilized in the previous section to the collapsed  gaze and gaze object representations. We note the latter two approaches use all timesteps in question for their prediction, and therefore are only comparable to the final timestep prediction. In the cumulative F1 approach (bottom graphs), we see that the time series approach matches or outperforms the other approaches. However, the collapsed eye gaze approach performs nearly as well using a simpler model. In contrast, the gaze object approach of  \cite{wachowiak}, which collapses the spatial dimension and only uses the frequency at which humans look at different objects, drastically reduces performance. These results indicate that the spatial dimension of  gaze data is more useful for predicting proficiency and trust compared to the temporal dimension.

\textbf{RQ3: Task time}
We now examine how predictive power changes as humans move from starting a new tasks to a phase of continuous execution of the task. \cref{fig:results_200} show the predictions curves when we start predicting at timestep $200$. Compared to the previous results, there are two trends. First, gameplay data is now strongly predictive from the first observed timestep, indicating that the state space itself contains a significant amount of information about the quality of play up to that point. Second, we see that the per-timestep prediction of the ungrounded gaze model no longer has the spike in prediction accuracy at the beginning, but rather consistently predicts trust and proficiency at a similar level to its predictions around timesteps $20$. Notably, even without this beneficial early bump, the cumulative prediction accuracy (bottom graphs) increases over time and achieves a substantially higher F1 score than any single timestep, indicating that even with no grounding, repeated measures of eye gaze data contain a rich signals about human behavior.

\section{DISCUSSION AND CONCLUSION}
In this paper, we collect a large dataset of human gameplay and gaze data while collaborating with a variety of different agents, specifically a random agent, a self-play(SP) agent and a HAHA \cite{haha}, in the fast-paced simulated environment of ``Overcooked." We then use this dataset to thoroughly examine the predictive ability of various implicit human signals.
We highlight the following findings:
1) Both eye gaze data and gameplay data provide strong predictive signals for human proficiency, their trust in an autonomous agent, and their intent; 
2) eye gaze data, even when not grounded in the environment, provides its predictive power early on, and is superior to gameplay data at the start of tasks when humans are deciding how to act and few actions have been performed. As more human behavior is observed, gameplay data catches up and eventually surpasses ungrounded eye gaze data;
3) Combining both eye gaze data and gameplay data provides the best overall predictive ability;
4) Caution should be applied when aggregating eye gaze data. If eye gaze data is to be aggregated, our findings support aggregation over the temporal dimension as preferable over aggregation over the spatial dimension.
We note that while in our experiments temporal aggregation only had a minimal impact, certain tasks or domains may be more sensitive to it.

\subsection{Future Work} 

Our findings underpin two key future research directions.
First, we are interested in investigating the potential for enhancing the adaptability and personalization of autonomous agents by conditioning them on the information collected about their human teammates.
Second, we intend to apply and extend these findings to real-world human-robot collaboration. While we are confident that our general conclusions will extend to these practical scenarios, a real-world domain raises number of interesting questions.
These include how to adapt the systems to account for the movement of the human, how to classify the completion of a human action, and determining an appropriate frequency to use when delineating timesteps.
Additionally, one potential limitation of this study is that our data was drawn from a participant pool relatively lacking in terms of age, ethnic and cultural diversity. Considering evidence for the culturally contingent nature of gaze patterns (e.g., \cite{zhang2021cross}), future work should explore the cultural nuances of eye gaze as a communicative signal. This is particularly relevant in diverse and multicultural settings where human-robot interactions may be influenced by varying interpretations of gaze behavior. 

\section*{ACKNOWLEDGMENT}
This work was supported by the Army Research Laboratory under Grants W911NF-21-2-02905
and W911NF-21-2-0126 and by the Office of Naval Research under Grant N00014-22-1-2482. The authors thank Abdul Ariyo, Zoe Iwu, Anna Madison, and R.J. Long for their assistance.

\bibliography{IEEEexample}
\bibliographystyle{IEEEtran}




\end{document}